\documentclass[11pt]{article}

\makeatletter
\newif\if@restonecol
\makeatother

\usepackage{vmargin}
\usepackage{comment}
\usepackage[ruled]{algorithm2e}
\usepackage{url}
\usepackage{graphicx,subfigure}
\usepackage{amsmath,amssymb,latexsym} 
\usepackage{multirow,paralist}
\usepackage{wrapfig}
\usepackage[breaklinks=true,pdfstartview=FitH]{hyperref}

\setpapersize{USletter}
\setmarginsrb{1in}{1.5in}{1in}{0.5in}{0in}{0in}{11pt}{0.4in} 

\newlength{\defbaselineskip}
\begin{document}

\title{Multi-Scale Link Prediction%
}

\author{
Donghyuk Shin
\thanks{
Dept. of Computer Science,
University of Texas at Austin,
Austin, TX 78712, USA
Email: dshin@cs.utexas.edu
}
\and
Si Si
\thanks{
Dept. of Computer Science,
University of Texas at Austin,
Austin, TX 78712, USA
Email: ssi@cs.utexas.edu
}
\and
Inderjit S. Dhillon
\thanks{
Dept. of Computer Science,
University of Texas at Austin,
Austin, TX 78712, USA
Email: inderjit@cs.utexas.edu
}
}

\date{}
\maketitle

\maketitle
\begin{abstract}
The automated analysis of social networks has become an important problem due to the proliferation of social networks, 
such as LiveJournal, Flickr and Facebook. The scale of these social 
networks is massive and continues to grow rapidly. An important problem in social network analysis is 
proximity estimation that infers the closeness of different users. Link prediction, in turn, is an important 
application of proximity estimation. However, many methods 
for computing proximity measures have high computational complexity and are thus prohibitive 
for large-scale link prediction problems. One way to address this problem is to estimate proximity measures via 
low-rank approximation. However, a single low-rank approximation may not be sufficient to represent the 
behavior of the entire network. In this paper, we propose Multi-Scale Link Prediction (MSLP), a 
framework for link prediction, which can handle massive networks. The basis idea of MSLP is 
to construct low rank approximations of the network at {\em multiple scales} in an efficient manner. 
Based on this approach, MSLP combines predictions at multiple scales to make robust and 
accurate predictions. Experimental results on real-life datasets with more than a million nodes show the superior performance and scalability of our method.
\end{abstract}

\vspace*{1ex}
\section{Introduction}
Social network analysis has become essential due to the proliferation of social networks, such as 
LiveJournal, MySpace, Flickr and Facebook. The scale of these social networks is massive and continues 
to grow rapidly. For example, Facebook now has more than 800 million active users with over 700,000 new users joining everyday. 
This has significantly changed the way people interact and share information with others and has led to unprecedented research opportunities.

An important problem for social network analysis is proximity estimation that 
infers the ``closeness'' of different users. Proximity measures quantify the 
interaction between users based on the structural properties of a graph, such 
as the number of common friends. A key application for proximity measure in 
social networks is link prediction, which is a key 
problem in social network analysis \cite{LibenNowell:2003bo,Song:2012sig}.
Applying proximity measure for link prediction is based on the assumption that 
a pair of users with a high proximity score indicates they are close in terms of 
social relatedness and thus this pair of users will have a good chance to 
become friends in the future.

Simple proximity measures, such as neighborhood-based measures, e.g., Adamic-Adar score
\cite{AA:03} and common neighbors \cite{Newman:2001CN}, can be computed efficiently. However, they describe a very localized view of interaction. There are more comprehensive proximity measures that capture a broader perspective of social 
relationships by considering all paths between users. These path based methods, 
such as Katz \cite{Katz:1953} or Rooted PageRank \cite{LibenNowell:2003bo}, are often more effective. 
Nonetheless, they are also well known for their high computational complexity and memory usage, which limits
their applicability to massive graphs with more than a million of users.

To solve this problem, a great deal of work has been done on scalable proximity estimation \cite{Bonchi:2011vc,Song:2009vv}. 
One basic idea is to perform dimensionality reduction on the original graph and then compute the proximity based on its low-rank approximation. Recently, clustered low rank approximation (CLRA) has been proposed which develops a fast and memory efficient 
method \cite{Savas:2011vl,Song:2012sig}.

However, a single low-rank approximation may not be sufficient to represent the whole network. Furthermore, the approach in \cite{Savas:2011vl} only uses a single clustering structure making it sensitive to a particular clustering and biased against links that happen to be between clusters. More notably, a recent study has shown that large social networks tend to lack large well-defined clusters which suggests that a single clustering structure can be problematic \cite{Leskovec:2009tm}.

To address the above problems, we propose a multi-scale approximation of the graph to obtain multiple granular views of the network in order to perform link prediction in a scalable and accurate manner. This is achieved by taking a hierarchical clustering approach and generating low-rank approximations at each level in the hierarchy. 
Although we use a single hierarchical representation of the graph, we do not require it to be the optimal structure. The main purpose of using hierarchical clustering is not 
to detect the underlying community structure of the graph, but to use
it as a tool for efficient multi-scale approximation.
In the experimental section, we show that under various clustering structures even in the case with poor clustering quality, our proposed algorithm can still achieve better results compared to other link prediction algorithms (for example, in the \texttt{Epinions} network \cite{Richardson03trustmanagement} which was also used in \cite{Leskovec:2009tm}).

Specifically, in this paper we propose a robust, flexible, and scalable framework for link prediction on social networks that we call, \emph{multi-scale link prediction} (MSLP). 
MSLP exploits different scales of low rank approximation of social networks by combining information from multiple levels in the hierarchy in an efficient manner.
Higher levels in the hierarchy present a more global view while lower levels focus on more localized information.
MSLP works by first performing hierarchical clustering on the graph by utilizing a fast graph clustering algorithm, and then performing multi-scale approximation based on the produced hierarchy.
Since different levels have different approximation, each level will give different approximated proximity scores.
MSLP will combine approximated proximity scores from each level and make the final prediction based on the combined scores. 
As a result, MSLP captures both local and global information of the network.

We list the benefits of our framework as follows:
\begin{itemize}
\item MSLP makes predictions based on information from multiple scales and thus can make more accurate and robust predictions.
\item MSLP is fast and memory-efficient as it uses a simple and fast tree-structured subspace approximation method, which speeds up the computation of our multi-scale approximation while re-using memory across different levels. 
As a result, it can be applied to social networks with millions of users.
\item MSLP is flexible in two aspects: (1) it can be used with any other reasonably good clustering algorithm to 
generate a multi-scale view of the graph as it does not depend on a particular hierarchical structure, (2) as a dimensionality reduction method, we are not tied down to a particular proximity measure, e.g., Katz and CN, and others can be used.
\end{itemize}

The rest of the paper is organized as follows. In Section 2, we survey some related work on link prediction. 
Next, some background material is introduced in Section 3. In Section 4, we propose our algorithm: MSLP. 
Experimental results on real world large-scale social networks are presented in Section 5. Finally, we present our conclusions in Section 6.

\section{Related Work}
Link prediction refers to the problem of inferring new interactions among members in a network. The first systematic treatment of the problem appeared in \cite{LibenNowell:2003bo}, where a variety of proximity measures, such as Common Neighbors \cite{Newman:2001CN} and the Katz measure \cite{Katz:1953} are used as effective methods for link prediction. 
In addition to unsupervised approaches, there is also rising interest in supervised approaches 
for link prediction \cite{Backstrom:2011bi,Hasan:06,Lichtenwalter:2010cf}. In supervised link prediction, node and/or edge features are extracted from the network and treated as a classification problem. 
However, engineering good features and how to encode the class imbalance problem are still challenging tasks. 
Recently, link prediction has been shown to benefit from exploring additional information external to the network, such as node or edge attributes \cite{Hsu07structurallink,Scellato:2011}. However, these approaches require additional information, which may be difficult to obtain due to privacy and security issues.

Many popular proximity measures that are used for link prediction have high computational complexity and 
do not scale well to large-scale networks. A great deal of recent work has been devoted to speedup the computation. 
For example, \cite{Wang:2007} truncates the series expansion of Katz and only considers paths of length up to some threshold.
In \cite{LibenNowell:2003bo,Song:2009vv}, dimensionality 
reduction methods, such as the eigen-decomposition, are used to construct low rank approximations 
of a graph, which are then used to compute approximated proximity measures. 
The more recent work in \cite{Bonchi:2011vc} applies the Lanczos/Stieltjes procedure 
to iteratively compute upper and lower bounds of a single Katz value and shows that these eventually converge to the real Katz value.

Efficient proximity estimation is essential for scalable link prediction. 
However, one should be able to make accurate and robust predictions with the 
estimated measures. For example, \cite{Song:2009vv,Song:2012sig} explore the low rank 
approximation of social networks to speed up large-scale link prediction. 
Another way to improve the link prediction performance is to explore the community structure of a network. 
For example, LinkBoost \cite{Comar:2011} explores the community structure by a novel degree dependent cost 
function and shows that minimization of the associated risk can lead to more links predicted within communities 
than between communities. However, considering a single community structure may not lead to robust predictions, 
because even detecting the `best' community structure itself is still an open 
question.

Very little work has been done using hierarchical structures for link prediction. One exception is the method proposed by \cite{Clauset:2008fxa}, which works by sampling a number of competitive 
hierarchical random graphs from a large pool of such graphs. Each sampled 
graph is associated with a probability indicating the strength of community 
structure over the original network. The probability of a link appearing 
between any two nodes is averaged over the corresponding connecting 
probability on the sampled graphs. However, to predict potential links, this 
algorithm needs to enumerate and average over almost all possible hierarchical 
partitions of a given network and thus is very costly to compute even with 
small networks. Compared with \cite{Clauset:2008fxa}, our 
algorithm is much more efficient in terms of speed and thus can be scaled up 
to large-scale link prediction problems with millions of users.

\section{Background}
Assume we are given a graph $G = (\mathcal{V},\mathcal{E})$, where $\mathcal{V}=\{1,\cdots,n\}$ is the set of vertices representing the users in a social network and $\mathcal{E}=\{e_{ij}|i,j\in\mathcal{V}\}$ is the set of edges quantifying the connection between user $i$ and user $j$. Let $A=[a_{ij}]$ 
be the corresponding $n\times n$ adjacency matrix of $G$ such that $a_{ij}=e_{ij}$, if there is an edge between $i$ and $j$ and 0 otherwise.
For simplicity, we assume $G$ is an undirected graph, i.e., $A$ is symmetric.

As shown in \cite{LibenNowell:2003bo}, proximity measures can be computed from $A$. Many of these measures can be represented as a matrix function $f(A)$, where the $(i,j)$-th element represents the value of a proximity measure between user $i$ and user $j$ \cite{Higham:2010SR}.
One popular measure is the common neighbor, which can be captured by $f_{cn}(A)=A^2$, describing a very localized view of interactions between vertices by considering only paths of length $2$. A more extensive measure is the popular Katz measure \cite{Katz:1953}. Such path-based proximity measures often achieve better accuracy at the cost of higher computational complexity. The Katz measure is defined as follows
\[
  f_{kz}(A) = \sum_{k=1}^{\infty}\beta^k A^k=(I-\beta A)^{-1} - I,
\]
where $I$ is the identity matrix and $\beta\leq1/{\|A\|_2}$ is a damping parameter. As we can see, both measures takes $O(n^3)$ time, which is computationally infeasible for large-scale networks with millions of nodes.

Here, dimensionality reduction methods, such as the singular value decomposition (SVD), play an important role. These methods are particularly useful, since it suffices to have a reasonably good estimation of a given proximity measure for most applications. Furthermore, low rank approximation of the adjacency matrix serves as a useful conceptual and computational tool for the graph.

Assume that we are given a rank-$r$ approximation of the $n\times n$ matrix $A$ as follows
\[
  A \approx \tilde{A} = USU^T,
\]
where $U$ is an $n\times r$ orthonomal matrix (i.e. $U^TU = I_r$ is an identity matrix), and $S$ is an $r\times r$ matrix. Using this low-rank approximation $\tilde{A}$, the CN measure can be approximated as
$f_{cn}(A) \approx US^{2}U^T$. Similarly, the Katz measure may be approximated by
\begin{align*}
   f_{kz}(A) \approx & \sum_{k=1}^{\infty}\beta^k \tilde{A}^k
   = \sum_{k=1}^{\infty}\beta^k(USU^T)^k\\
   =& U(\sum_{k=1}^{\infty}\beta^kS^k)U^T
   = U((I-\beta S)^{-1}-I)U^T,
\end{align*}
respectively. In general, $f(A) \approx Uf(S)U^T$, which requires less computational resources as the matrix function is only evaluated on the small $S$ matrix.

However, computing the low rank approximation of a massive graph via SVD or 
other popular dimensionality reduction methods can still be a computational 
bottleneck. Recently, the technique of Cluster Low Rank Approximation (CLRA) was proposed by 
\cite{Savas:2011vl} as a scalable and accurate low rank approximation method. The 
basic idea of CLRA is to preserve important structural information by 
clustering the graph $G$ into $c$ disjoint clusters. Then it computes a low 
rank approximation of each cluster, which is extended to approximate the 
entire graph as a final step.

Assume that the graph has been clustered into $c$ clusters and the vertices are ordered as follows
\[
A =
\begin{bmatrix}
A_{11} & \cdots & A_{1c}\\
\vdots & \ddots & \vdots\\
A_{c1} & \cdots & A_{cc}
\end{bmatrix},
\]
where the diagonal blocks $A_{ii}$, $i=1,\ldots,c$, correspond to the local adjacency matrix of each cluster $i$. For every cluster, the best rank-$r$ approximation is computed as
$
A_{ii}\approx U_i\Lambda_iU_i^T,
$
where $\Lambda_i$ is a diagonal matrix with the $r$ largest eigenvalues of $A_{ii}$, and $U_i$ is an orthonomal matrix with the corresponding eigenvectors. Finally, CLRA aligns the low rank approximations of each cluster together to obtain the clustered low rank approximation of the entire adjacency matrix $A$. Mathematically,
\[
A \approx 
\begin{bmatrix}
U_{1} & \ldots & 0\\
\vdots & \ddots & \vdots\\
0 & \ldots & U_{c}
\end{bmatrix}
\begin{bmatrix}
S_{11} & \ldots & S_{1c}\\
\vdots & \ddots & \vdots \\
S_{c1}& \ldots & S_{cc}
\end{bmatrix}
\begin{bmatrix}
U_{1} & \ldots & 0\\
\vdots & \ddots & \vdots\\
0 & \ldots & U_{c}
\end{bmatrix}^T,
\]
where $S_{ij}=U_{i}^{T}A_{ij}U_{i}$, for $i,j=1,\ldots,c$, which is the optimal $S$ in the least squares sense. Note that the block-diagonal matrix $U = \text{diag}(U_1,\ldots,U_c)$ is also orthonomal and $S_{ii}=\Lambda_i$ are diagonal. It is shown in \cite{Savas:2011vl} that CLRA achieves accurate approximations while being efficient in both computational speed and memory usage.
However, the drawback of CLRA is that it only uses one clustering structure, whereas it has been shown that many large social networks lack such structure \cite{Leskovec:2009tm}. In this paper, we overcome such limitation by taking a multi-scale approach.

Based on the estimated proximity measures, we can perform link prediction on 
social networks. Link prediction deals with the evolving networks with time stamp 
information. Specifically, given a snapshot of a network $A_{t_1}$ at time 
$t_1$, the task is to predict links that would form in $A_{t_2}$ at a future 
time step $t_2$. A high proximity score between two users implicitly states the high correlation 
between them and thus a high chance to form a new link in the future.

\section{Proposed Algorithm}
In this section we present our \emph{multi-scale link prediction} (MSLP) framework for social networks. Our method mainly consists of three phases: hierarchical clustering, subspace approximation and multi-scale prediction. Specifically, we first construct a hierarchy tree with a fast top-down hierarchical clustering approach. 
Then, a multi-scale low rank approximation to the original graph is computed when traversing the hierarchy in a bottom-up fashion. 
An important technical contribution of our paper is a fast tree-structured approximation algorithm that enables
us to compute the subspace of a parent cluster quickly by using subspaces of its child clusters; this allows us
to compute each level's low-rank approximation efficiently.
Finally, we combine proximity measures, which are computed using the multi-scale low rank approximation of the graph, and make our final predictions.

\vspace*{-2mm}
\subsection{Hierarchical clustering}
The first step of our method is to hierarchically cluster or partition a given graph. The purpose of this is to efficiently generate a multi-scale approximation of the graph using the constructed hierarchical structure. This, in turn, makes predictions more accurate and robust as we combine predictions at each level of the hierarchy in the final step.

Generally, there are two main approaches for hierarchical clustering: 
agglomerative (or bottom-up) approach and divisive (or top-down) approach. 
The agglomerative approach initially treats each vertex as one cluster and continually merges pairs of clusters as it moves up the hierarchy. The divisive approach takes the opposite direction, that is, all vertices are placed in a single cluster 
and recursively partitioned into smaller clusters. Due to the large scale of the problem and the availability of 
efficient clustering software, such as Graclus \cite{Dhillon:2007kp}, we employ the divisive approach in our work.

Given a graph $G=(\mathcal{V},\mathcal{E})$, our goal is to construct a level $\ell$ hierarchy, so as to generate a multi-scale view of the graph. We form $c$ nodes at the first level of the hierarchy by clustering $\mathcal{V}$ into $c$ disjoint sets $\mathcal{V}_{1}^{(1)},\mathcal{V}_{2}^{(1)},\ldots,\mathcal{V}_{c}^{(1)}$, 
where the superscript denotes the level of the hierarchy. Then, we proceed to the second level of the hierarchy tree by further 
clustering each node in the first level $\mathcal{V}_{i}^{(1)}$, $i=1,\ldots,c$, and generate $c$ child nodes from each of them as  
$\mathcal{V}_{i1}^{(2)},\mathcal{V}_{i2}^{(2)},\ldots,\mathcal{V}_{ic}^{(2)}$, such that $\mathcal{V}_{i}^{(1)}=\bigcup_{j=1}^c \mathcal{V}_{ij}^{(2)}$. 
As a consequence, we will have $c^2$ nodes on the second level of the hierarchy. This process repeats until the desired number of levels $\ell$ is reached. Many classic clustering methods can be used as a base clustering method. In this paper, we use the Graclus algorithm \cite{Dhillon:2007kp} to cluster each node because of its ability to scale up to very large graphs. However, our 
algorithm can be combined with any other graph clustering method.

  \begin{table}[b!]
  \vspace*{-1em}
 	\caption{Percentage of within-cluster edges using Graclus. Numbers in brackets represent random clustering. It can be seen that Graclus is quite effective in finding good clustering structure. (these networks contain about 2 million nodes --- details are given in Table \ref{table:dataset2})}
	\label{table:hier_clust_stat}
	\centering
	\begin{tabular}{c||c|c|c}\hline
	Hierarchy & \texttt{Flickr} & \texttt{LiveJournal} & \texttt{MySpace} \\
	\hline\hline
	Level 1 & 96.2 (68.1) & 99.3 (60.1) & 98.6 (61.6) \\
	Level 2 & 95.1 (61.7) & 98.8 (51.7) & 88.0 (35.2) \\
	Level 3 & 88.1 (54.3) & 85.0 (28.6) & 69.5 (18.0) \\
	Level 4 & 85.2 (51.4) & 79.4 (15.2) & 64.3 (13.0) \\ 
	Level 5 & 66.7 (27.2) & 70.0 (9.4) & 56.3 (8.4) \\
	\hline
	\end{tabular}
	\vspace*{-3mm}
  \end{table}

The hierarchy of $A$ at level $p$, after sorting the vertices, can be written as
  \[A = \begin{bmatrix}
	A_{11}^{(p)} &\ldots& A_{1\hat{c}}^{(p)} \\
	\vdots &\ddots& \vdots \\
	A_{\hat{c}1}^{(p)} &\ldots& A_{\hat{c}\hat{c}}^{(p)}
  \end{bmatrix},\]
where $\hat{c}=c^p$ is the number of nodes in level $p$ and each diagonal block 
$A_{ii}^{(p)}$, $i =\begin{tiny}•\end{tiny} 1,\ldots,\hat{c}$, is an $m_i\times m_i$ matrix that can be viewed as a local adjacency matrix of cluster $i$ at level $p$. 
The off-diagonal $m_i\times m_j$ blocks, $A_{ij}^{(p)}$, where $i\neq j$, contains the set of edges between clusters $i$ and $j$. 

As it is desirable to capture most of the links within clusters, we compare with random clustering in terms of the percentage of within-cluster links on three large-scale social networks in Table \ref{table:hier_clust_stat}. 
For each level of the hierarchy tree, the within-cluster links are those that connect two nodes in the same cluster. As shown in Table \ref{table:hier_clust_stat}, the percentage of within-cluster edges of random clustering 
is much smaller than the hierarchical clustering scheme used in this paper, and the gap becomes much 
larger when going down the hierarchy. Even at the deepest level, the 
clustering scheme we use can still capture more than half of the edges 
compared with less than 10\% in the \texttt{LiveJournal} and \texttt{MySpace} graphs when using random clustering.

As a final remark, we note that the hierarchical clustering scheme is also very fast. Clustering the above three networks into 5 levels with 2 clusters at each level can be completed in three hundred seconds on a 8-core 3.40GHz machine. 
In the next section, we show how to use the hierarchy structure to construct a multi-scale 
approximation of large-scale graphs efficiently. 

\vspace*{-3mm}
\subsection{Subspace approximation}
\begin{figure}[h!]
	\centering
	\includegraphics[scale=0.6]{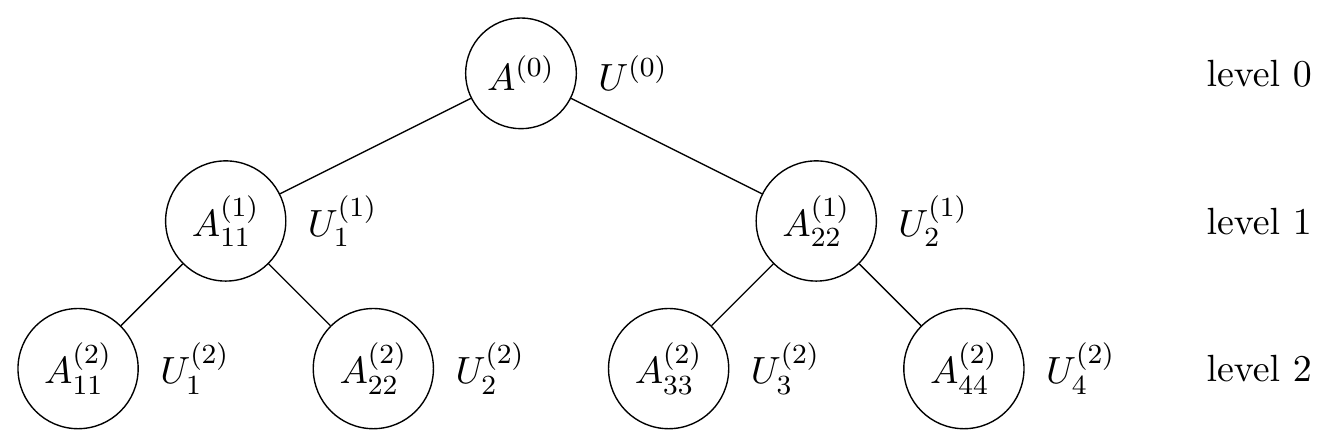}
	\caption{Hierarchical structure example.}
	\label{fig:example}
	\vspace*{-4mm}
  \end{figure}

After constructing the hierarchy for a given graph, we can compute low rank approximations of $A$ at each level of the hierarchy to obtain a multi-scale approximation. Specifically, we employ CLRA to obtain the approximation. 
Figure \ref{fig:example} gives an example of a simple three level hierarchy to better illustrate our method. By applying CLRA on each of the 3 levels in the example, we have 3 clustered low rank approximations of $A$ as follows

\begin{itemize}
\item Level 0:
\[
A \approx \tilde{A}^{(0)} = U^{(0)}S^{(0)}U^{(0)^T},
\]
\item Level 1:
\[
A \approx \tilde{A}^{(1)} =
\begin{bmatrix}
U_{1}^{(1)} & 0\\
0 & U_{2}^{(1)}
\end{bmatrix}
\begin{bmatrix}
S_{11}^{(1)} & S_{12}^{(1)}\\
S_{21}^{(1)} & S_{22}^{(1)}
\end{bmatrix}
\begin{bmatrix}
U_{1}^{(1)} & 0\\
0 & U_{2}^{(1)}
\end{bmatrix}^T,
\]
\item Level 2:
{\setlength\arraycolsep{0.1em}
\begin{eqnarray*}
  &&A \approx \tilde{A}^{(2)} = U^{(2)}S^{(2)}U^{(2)^T} \\
&&=
\begin{bmatrix}
U_{1}^{(2)} & \ldots & 0\\
\vdots & \ddots & \vdots\\
0 & \ldots & U_{4}^{(2)}
\end{bmatrix}
\begin{bmatrix}
S_{11}^{(2)} & \ldots & S_{14}^{(2)}\\
\vdots & \ddots & \vdots \\
S_{41}^{(2)} & \ldots & S_{44}^{(2)}
\end{bmatrix}
\begin{bmatrix}
U_{1}^{(2)} & \ldots & 0\\
\vdots & \ddots & \vdots\\
0 & \ldots & U_{4}^{(2)}
\end{bmatrix}^T,
\end{eqnarray*}}
\end{itemize}
where $U_{i}^{(p)}$ is the set of orthonormal basis forming the subspace for cluster $i$ on level $p$ and $S_{ij}^{(p)}={U_{i}^{(p)^T}}A_{ij}^{(p)}U_{j}^{(p)}$, $1\leq i,j\leq 2^p$. 
Level 0 can be viewed as a special case of CLRA, where the entire graph is treated as a single cluster, which yields a global view of the entire matrix $A$. Lower levels in the hierarchy will preserve more local information within each cluster. Thus, each level of approximation concentrates on different levels of granularity, resulting in a multi-scale approximation of $A$.

\begin{figure}[t!]
  \vspace*{-3mm}
	\centering
	\includegraphics[scale=0.8]{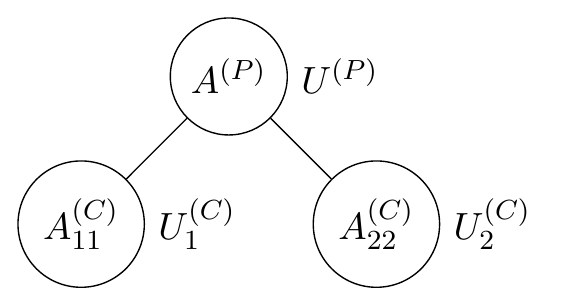}
	\caption{Approximating parent cluster's subspace.}
	\label{fig:basis}
	\vspace*{-3mm}
  \end{figure}

An important issue here is how to compute each level's approximation of $A$ efficiently. A straightforward solution would be use standard dimensionality reduction methods, such as SVD. 
This can be computed efficiently for clusters at the deepest level of the hierarchy tree, since the size of each cluster is relatively small. However, the computational cost becomes prohibitive as the size 
of the cluster increases, which is the case for upper levels in the hierarchy tree. 
We propose a more scalable and effective method to address this issue. For clarity and brevity, we focus on a local view of the hierarchy as shown in Figure \ref{fig:basis}, where $A^{(P)}$ is a parent cluster and $A_{11}^{(C)}$ and $A_{22}^{(C)}$ are its two child clusters.

A key observation we make is that the subspaces between any two adjacent levels in the hierarchy tree 
should be close to each other. That is, $U^{(P)}$ should be close to $\text{diag}(U_1^{(C)},U_2^{(C)})$. This is because many of the links that are in $A^{(P)}$ should be captured by its child clusters $A_{11}^{(C)}$ and $A_{22}^{(C)}$. Thus, if we are given $\text{diag}(U_1^{(C)},U_2^{(C)})$, one should be able to compute $U^{(P)}$ faster than computing it from scratch. Thus, we propose an algorithm that uses the child cluster's subspace to compute the parent cluster's subspace.

Our proposed method, \emph{tree-structured approximation of subspace}, is listed in Algorithm \ref{alg:alg1}. 
The main idea is to construct a matrix $Y=A^{(P)}\Omega$ that covers as much of the range space of $A^{(P)}$ as possible. This can be done efficiently using $\Omega=\text{diag}(U_{1}^{(C)},U_{2}^{(C)},\ldots,U_{c}^{(C)})$. Note that both $A^{(P)}$ and $\Omega$ are sparse. Then, an orthonormal matrix $Q$ is computed from $Y$ as a basis for the range of $Y$ (e.g. using the QR-decomposition). 
Finally, $Q$ is further used to compute $U^{(P)}$ via standard factorizations, such as eigen-decomposition, on the matrix $B=Q^TAQ$. This last step is also fast since $B$ is a small $cr\times cr$ matrix, where $r$ is the rank of the approximation of each child node.

  \begin{algorithm}[b!]
  \DontPrintSemicolon
	\KwIn{$n\times n$ adjacency matrix of parent cluster $A=A^{(P)}$, child cluster's subspaces $U_{1}^{(C)},\ldots,U_{c}^{(C)}$, target rank $r$.}
	\KwOut{dominant subspace for parent cluster $A^{(P)}$, i.e. $U^{(P)}$.}
	\BlankLine
	$\Omega \leftarrow \text{diag}(U_{1}^{(C)},U_{2}^{(C)},\ldots,U_{c}^{(C)})$.\\
	Compute {$n\times cr$} matrix $Y=A\Omega$.\\
	Compute $Q$ as an orthonormal basis for the range of $Y$.\\
	Compute $B=Q^TAQ$.\tcp*[r]{$A\approx Q(Q^TAQ)Q^T$}
	Compute rank-$r$ eigen-decomposition of  $B\approx V\Lambda V^T$.\\
	Compute $U^{(P)}=QV$.
	\caption{Tree-structured approximation of dominant subspace of parent cluster from child clusters}
	\label{alg:alg1}
  \end{algorithm}

The subspace approximation scheme in Algorithm \ref{alg:alg1} is more efficient than truncated eigen-decomposition (EIG), since the latter needs to be computed from scratch and is time consuming when dealing with large-scale matrices. We note that $Y=(AA^T)A\Omega$ can be used for higher accuracy, though we did not find any significant improvement in the results.
   
\begin{figure}[h!]
 	\centering
 	\includegraphics[scale=0.35]{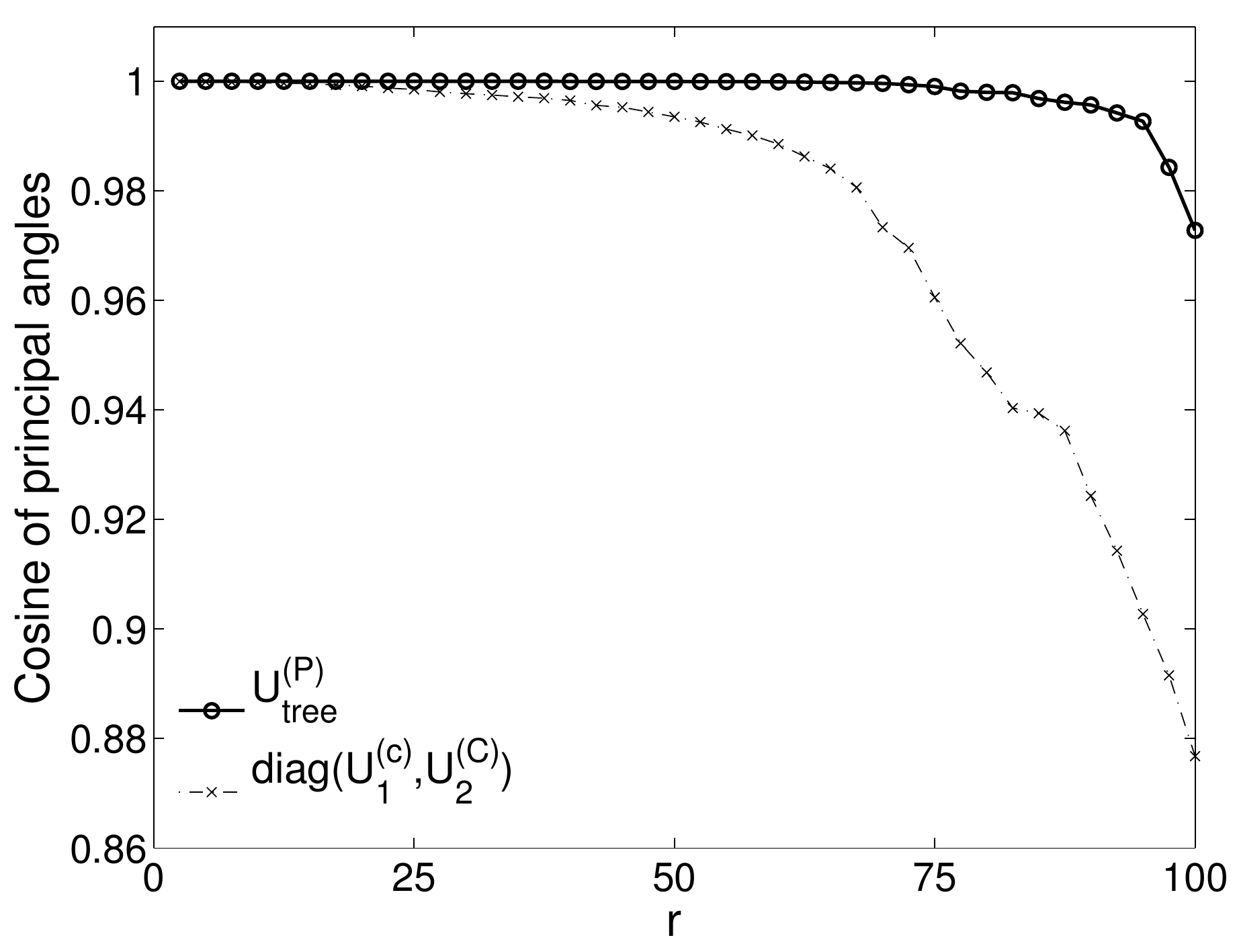}
 	 	\vspace*{-4mm}
 	\caption{Principal angles between parent cluster's subspace and two other subspaces: child cluster's subspace $\text{diag}(U_1^{(C)},U_2^{(C)})$ and parent cluster's subspace $U_{tree}^{(P)}$ computed using Algorithm \ref{alg:alg1}.}
 	\label{fig:pa}
   \end{figure}

Figure \ref{fig:pa} shows principal angles between the parent cluster's subspace $U_{eig}^{(P)}$ computed via eigen-decomposition and the child cluster's subspace $\text{diag}(U_1^{(C)},U_2^{(C)})$ for the \texttt{Flickr} dataset. The cosine of principal angles are close to $1$, supporting our observation that the subspaces of two adjacent levels are close to each other. We also show principal angles between $U_{eig}^{(P)}$ and the parent cluster's subspace $U_{tree}^{(P)}$ computed using Algorithm \ref{alg:alg1}. We see that these subspaces are even more closer to each other, showing that our algorithm can accurately approximate the parent cluster's subspace.

\vspace*{-4mm}
\subsection{Multi-scale link prediction}
As mentioned in Section 2, many proximity measures for link prediction $f(A)$ are expensive to compute on large-scale networks because of their high complexity. One solution is to approximate $A$ by a low-rank approximation $\tilde A$ and then compute approximated proximity measures with $f(\tilde A)$ to make predictions. This stems from the idea that most of the action of $A$ can be captured by a few latent factors, which can be extracted with low rank approximations of $A$.

\begin{algorithm}[t!]
\DontPrintSemicolon\SetNoFillComment
\KwIn{adjacency matrix $A$, number of levels $\ell$, number of clusters $c$ at each node, target rank $r$, weights $w_0,w_1,\ldots,w_\ell$.}
\KwOut{top-$k$ predictions.}
\BlankLine
\tcc{Hierarchical clustering}
$A_{11}^{(0)} \leftarrow A$.\\
\For{$i=0$ \KwTo $\ell$}{
	\For{$j=1$ \KwTo $c^i$} {
		Cluster $A_{jj}^{(i)}$ into $c$ clusters.\tcp*[r]{e.g. Graclus}
	}
}
\tcc{Subspace approximation}
\tcp{approximation for deepest level.}
Compute $U^{(\ell)}$, $S^{(\ell)}$ using CLRA.\\
\tcp{approximation for intermediate levels.}
\For{$i=\ell-1$ \KwTo $0$} {
	Compute $U^{(i)}$ using Algorithm \ref{alg:alg1}.\\
	$S^{(i)} = {U^{(i)}}^TAU^{(i)}$.
}
\tcc{Multi-scale prediction}
\For{$i=\ell$ \KwTo $0$} {
	$K_i = f(A^{(i)}) = U^{(i)}f(S^{(i)}){U^{(i)}}^T$.\tcp*[r]{e.g. Katz}
}
$P = w_0K_0+w_1K_1+\ldots+w_{\ell}K_{\ell}$.\\
\Return top-$k$ predictions according to $P$.
\caption{Multi-Scale Link Prediction (MSLP)}
\label{alg:alg2}
\end{algorithm}

It has been shown that CLRA provides an accurate and scalable low rank approximation, and can be used for efficient proximity estimation \cite{Song:2012sig}. However, CLRA just uses one clustering structure making it sensitive to a particular clustering and biased against links that appear between clusters.

Our proposed method alleviates such problem with a multi-scale approach. The main idea is that, under a hierarchical clustering, all links will eventually belong to at least one cluster. That is, even if we miss a between-cluster link at a certain level, it still has a good chance of getting corrected by upper levels as it will eventually become a within-cluster link. Moreover, links that lie within clusters at multiple levels, such as from the deepest level, gets emphasized multiple times. Those links will have the propensity of being included in the final prediction, which aligns with the intuition that links are more likely to form within tight clusters.

Once the multi-scale low rank approximation of $A$ is obtained, we now perform multi-scale link prediction. From each low rank approximation of the hierarchy, $\tilde{A}^{(i)}$, $i=0,1,\ldots,\ell$, the approximated proximity measure can be computed with $f(\tilde{A}^{(i)})$. This gives a total of $\ell+1$ proximity measures for each link, which are combined to make final predictions. Formally, our multi-scale predictions are given by
\begin{equation}
g( w_0f(\tilde{A}^{(0)}) + w_1f(\tilde{A}^{(1)}) + \ldots + w_lf(\tilde{A}^{(\ell)}) ), \nonumber
\end{equation}
where $w_i$'s are the weights for different levels and $g(\cdot)$ is the predictor, e.g. top-$k$ scoring links. For simplicity, we use the same weight for all levels in this work, i.e. $w_i=\frac{1}{\ell}$.

The entire flow of our proposed method, Multi-Scale Link Prediction (MSLP), is 
listed in Algorithm \ref{alg:alg2}. Next, we analyze the computation time and memory usage of MSLP.

\vspace{1ex}
\noindent {\bf Computation time:} 
As mentioned earlier, the hierarchical clustering is linear in the number of edges in the network and can be finished in a few hundred seconds on networks with 2 million nodes. Computing the approximated proximity scores as a final step for a given user is simply a matrix multiplication of low rank matrices and time complexity is 
$O(\ell nr^2)$. In general, we set the number of clusters $c$ and the rank in each 
cluster $r$ to be very small. Among the three phases of MSLP, the subspace approximation phase is the 
dominant part of the computation time.   
In Table \ref{table:time_stat}, we compare the CPU time for subspace approximation 
by Algorithm \ref{alg:alg1} and EIG on three large-scale social networks with about 2 million users. 
We can see that for each intermediate level from 4 to 0, the subspace approximation 
in MSLP is up to 10 times faster than that of EIG, demonstrating the effectiveness of
Algorithm \ref{alg:alg1}. 
Furthermore, since we operate on each cluster independently, MSLP can be easily parallelized to gain greater speedups.

 \begin{table}[t!]
    \caption{Computational time (in minutes) for subspace approximation by MSLP (Algorithm \ref{alg:alg1}) and EIG 
  on three large-scale social networks.}
	\label{table:time_stat}
	\centering
	\begin{tabular}{c|c||r|r|r}\hline
	\multicolumn{2}{c||}{Network} & \texttt{LiveJournal} & \texttt{Flickr} & \texttt{MySpace} \\
	\hline\hline
  	\multicolumn{2}{c||}{EIG}    & 157.10 & 146.28 & 211.29 \\ \hline
	\multirow{6}{*}{MSLP} & Level 0 & 30.74 & 29.98 & 38.27 \\ 
	& Level 1 & 20.18 & 21.95 & 35.86 \\
	& Level 2 & 18.93 & 17.25 & 29.17 \\
	& Level 3 & 14.01 & 15.33 & 26.36 \\
	& Level 4 & 13.74 & 15.79 & 26.36 \\
  	& Level 5 & 121.26& 132.93 & 188.26 \\ \cline{1-5}
	\multicolumn{2}{c||}{MSLP Total} & 218.88 & 233.33 & 344.02 \\
	\hline
	\end{tabular}
	\vspace*{-3mm}
  \end{table}

\vspace{1ex}
\noindent {\bf Memory usage:}
For a rank-$r$ approximation, EIG needs to store $r$ eigenvectors and eigenvalues which takes $O(nr+r)$ memory. 
Compared with EIG, CLRA is memory efficient as it only takes $O(nr+c^{2\ell}r^2)$ memory for a larger rank-$c^{\ell}r$ approximation \cite{Savas:2011vl}. 
MSLP basically has the same memory usage as CLRA. 
While MSLP achieves a multi-scale approximation, it is not necessary to store the subspaces for all levels simultaneously. We can reuse the memory allocated for the child cluster's subspace to store the parent cluster's subspace using Algorithm \ref{alg:alg1}.

\section{Experimental Evaluation}\label{experiment_section}
In this section we present experimental results that evaluate both accuracy and scalability 
of our method, Multi-Scale Link Prediction (MSLP), for link prediction. 
First we present a detailed analysis of our method using the Karate club network as a case study. 
This will give a better understanding of our algorithm and illustrate where it succeeds. Next we provide results under different parameter settings on a large social network. Lastly, we compare MSLP to other popular methods on massive real-world social networks with millions of users and demonstrate its good performance.

\subsection{Case study: Karate club}
\begin{figure}[t!]
\centering
\includegraphics[scale=0.28]{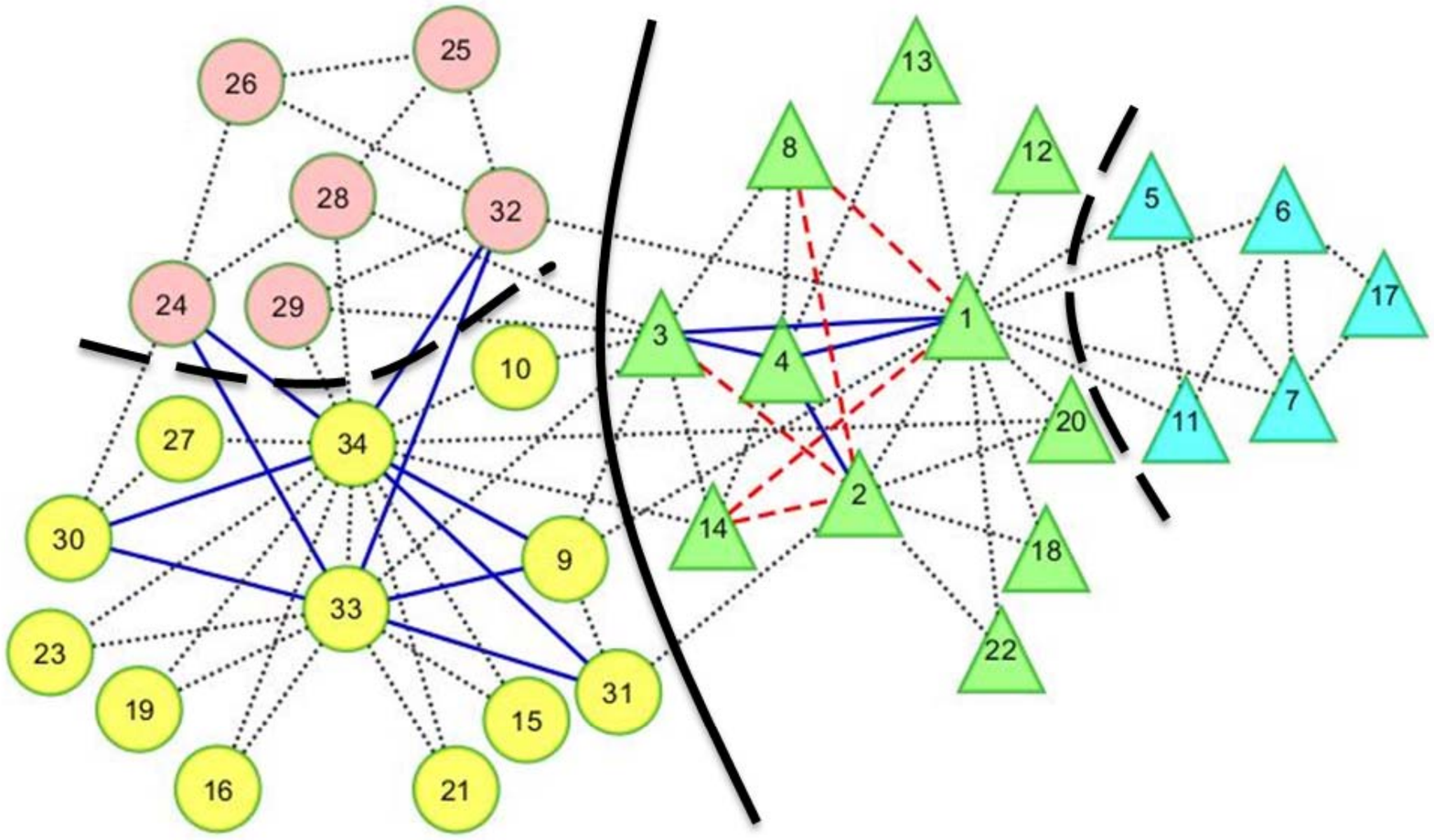}
\vspace*{-5mm}
\caption{Hierarchy of the Karate club network with 2 levels, two clusters on the first level and four clusters on the second level.}
\label{fig:karate2}
\vspace*{-4mm}
\end{figure}

We first start our performance analysis on a well-known small social network, Zachary's Karate club network \cite{Zachary:1977JAR}. 
The Karate club network represents a friendship network among 34 members of the club with 78 links. The clustering structure of the Karate club 
network is a standard example for testing clustering algorithms. We adopt the clustering results from 
\cite{refId0}, where the clustering is found via modularity optimization. 
Figure \ref{fig:karate2} shows the hierarchy of the Karate club network.
The first level has 2 clusters (circle and triangle) with 68 within-cluster links and the second level has 4 clusters (red, yellow, green and blue) with 50 within-cluster links. 	

As the Karate club network is a small network, we apply the 
\emph{leave-one-out} method to compare different methods. We first remove a 
single link from the network, treat the held out edge as 0, and perform link prediction 
on the resulting network. For each leave-one-out experiment, we compute the rank of the removed link based on its proximity measure. If the rank of the removed link appears in the top-$k$ list, we count it as a \emph{hit}. The number of top-$k$ hits is the number of hits out of all 78 links.

\begin{figure}[b!]
\vspace*{-3mm}
\centering
\includegraphics[scale=0.45]{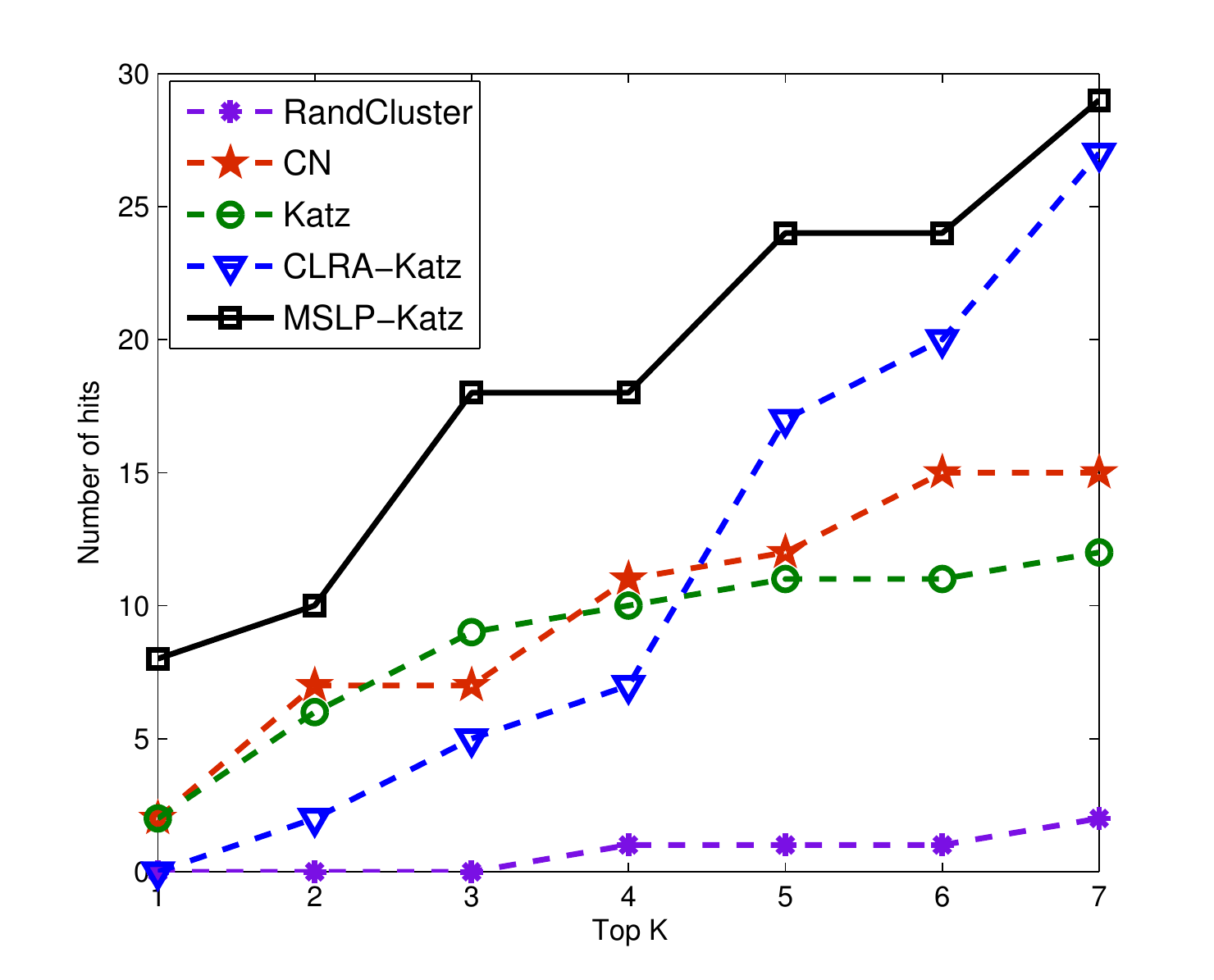}
\caption{Number of top-$k$ hits for different methods on the Karate club network.}
\label{fig:karate_results}
\vspace*{-3mm}
\end{figure}

We compare MSLP to four other methods: RandCluster, common neighbors (CN), 
Katz and CLRA. In RandCluster, we randomly partition the graph into 4 clusters 
and compute the Katz measure using CLRA with these clusters. Figure 
\ref{fig:karate_results} shows the number of top-$k$ hits for each method. 
Clearly, our method significantly outperforms other methods by achieving a 
much higher number of hits. This implies that MSLP makes more accurate 
predictions by considering the hierarchical structure of the network. 
RandCluster performs the worst, while CLRA has comparable performance with Katz indicating that the 
network's property can be captured by a few latent factors.

For a better illustration of the advantage of our method, we annotate Figure \ref{fig:karate2} with the results of top-3 hits. The solid blue links correspond to hits made by MSLP and the dashed red links are hits made by both CLRA and MSLP, i.e. the set of links successfully predicted by CLRA is a subset of that of MSLP. We can see that all hits made by CLRA are within-cluster links (green cluster), showing that CLRA favors within-cluster links. In contrast, MSLP can predict not only more within-cluster links, but also links between clusters (red and yellow). The ability to correctly predict both within and between-cluster links is one of the main advantages of our multi-scale approach.

\subsection{Experiments on Large Datasets}
In this section we present the results of link prediction on large real world datasets. We start by examining how the parameters of MSLP affects the performance. Particularly, we investigate how different hierarchical cluster structures impact the performance of MSLP. For this, we use a large real-world network: \texttt{Epinions}, which is an online social network from Epinions.com with 32,223 nodes and 684,026 links \cite{Richardson03trustmanagement}.

Next we use three real-world massive online social networks with millions of nodes: \texttt{Flickr} \cite{Mislove:2008ff}, \texttt{LiveJournal} and \texttt{MySpace} \cite{Song:2009vv}, and compare MSLP to other methods. These datasets have timestamps with them and we summarize each snapshot in Table \ref{table:dataset2}. The adjacency matrix at the first timestamp, $A_{t_1}$, is used to compute proximity measures, and the adjacency matrix at the next timestamp, $A_{t_2}$, is used for testing and evaluation. Since these networks are very large, we randomly select 5,000 users and evaluate on these users over all possible users. Performance measures are averaged over 30 iterations of such sampling. 

  \begin{table}[h!]
  \vspace*{-1em}
    \caption{Summary of networks with timestamps.}
    \label{table:dataset2}
	\centering
	 \begin{tabular}{l||r|r|r}	\hline
	  Network & Date & \# of nodes & \# of links \\
	  \hline\hline
	  \multirow{2}{*}{\texttt{Flickr}} & 5/6/2007 & 1,994,422 & 42,890,114\\ 
	  & 5/17/2007 & 1,994,422 & 43,681,874 \\ \hline
	  \multirow{2}{*}{\texttt{LiveJournal}} & 3/4/2009 & 1,757,326 & 84,366,676 \\
	  & 4/3/2009 & 1,757,326 & 85,666,494 \\ \hline
	  \multirow{2}{*}{\texttt{MySpace}} & 1/11/2009 & 2,086,141 & 90,918,158 \\
	  & 2/14/2009 & 2,086,141 & 91,587,516 \\
	  \hline\end{tabular}
  \end{table}
 
As pointed out in \cite{Leskovec:2008}, most of all newly formed links in 
social networks close a path of length two and form a triangle, i.e., appear 
in a user's 2-hop neighborhood. All three datasets show that this is the case 
for at least 90\% of test links in the second timestamp. For similar reasons 
in \cite{Backstrom:2011bi}, we focus on predicting links to users that are within its
2-hop neighborhood.

\subsubsection{Evaluation methodology}
We evaluate the accuracy of different methods by computing the \emph{true positive rat}e (TPR) and the \emph{false positive rate} (FPR), defined by
\begin{align*}
	\text{TPR} =& \frac{\text{\# of correctly predicted links}}{\text{\# of actual links}}, \\
	\text{FPR} =& \frac{\text{\# of incorrectly predicted links}}{\text{\# of non-friend pairs}},
\end{align*}
for all links in a sampled test set. Our evaluation is based on \emph{receiver operating characteristic} (ROC) curve and its \emph{area under the ROC curve} (AUC) that present achievable TPR with respect to FPR. Predicting links with proximity measures involves some thresholding on the measures to produce top-$k$ predictions. The ROC curves captures the full spectrum of prediction performance by varying the decision threshold. 

However, in a practical sense, a user is recommended only a small number of top-$k$ predictions and the hope is that most of them are correct. Thus, we focus on the region of low FPR by plotting FPR along the $x$-axis in log-scale, since it reflects the quality of these top-$k$ links. In the same spirit, we also use the \emph{Precision at Top}-$k$, i.e., the number of correct predictions out of top-$k$ recommendations, as our evaluation metric.

\vspace{1ex}
\noindent {\bf Other methods for comparison:}
We have carefully chosen a variety of proximity measures to compare with: Preferential Attachment (PA), Adamic-Adar score (AA), Random Walk with Restarts (RWR), common neighbors (CN) and Katz \cite{LibenNowell:2003bo}.
The actual values of Katz quickly becomes difficult to compute as scale increases due to its high computational cost. Therefore, we employ the Lanczos method \cite{Bonchi:2011vc} for its speed and good approximation to the real Katz values.
We also consider a supervised machine learning method (LR) \cite{Hasan:06}. For the latter, we extracted five network-based features: paths of lengths 3, 4 and 5, CN, and AA. Using these features, a logistic regression model is trained over a sampled set of positive and negative links from 100,000 users as in \cite{Hasan:06}.

\subsubsection{Varying parameters}
Next we evaluate on different parameter settings by varying the three main parameters of MSLP: number of levels in the hierarchy $\ell$, number of clusters each node in the hierarchy $c$, and rank $r$. We fix $\ell=3$, $c=2$, and $r=20$ while changing one parameter at a time and measure AUC and precision at top-20. The \texttt{Epinions} network does not have time information, thus we randomly sample a number of links and treat them as test links in $A_{t_2}$. The sampling is performed such that about 90\% of test links appear in a user's 2-hop neighborhood.

We compare the performance of our method to two other low-rank approximation methods: eigen-decomposition (EIG), clustered low rank approximation at the deepest level in the hierarchy tree (CLRA).

\vspace{1ex}
\noindent {\bf Rank:}
Table \ref{table:varying_r} shows performance of the three low-rank approximation methods with different ranks. MSLP consistently performs better in terms of both AUC and Precision at Top-20 than the other two methods. The accuracy of CLRA deteriorates as the rank increases. This implies that the low-rank approximation of each cluster starts to accumulate noise at larger ranks. It is shown that low-rank approximation methods tend to perform best at an intermediate rank \cite{LibenNowell:2003bo}, which is also the case here with $r=20$. 

  \begin{table}[h!]
      \caption{Varying rank of approximation on \texttt{Epinions} dataset. MSLP consistently outperforms EIG and CLRA for different ranks in terms of AUC and precision at top-20.}
	  \label{table:varying_r}
	\centering
	  \begin{tabular}{c||r|r||r|r||r|r}	\hline
	  \multirow{2}{*}{$r$} & \multicolumn{2}{|c||}{EIG-Katz} & \multicolumn{2}{|c||}{CLRA-Katz} & \multicolumn{2}{|c}{MSLP-Katz} \\ \cline{2-7}
	  & AUC & Prec & AUC & Prec & AUC & Prec \\
	  \hline\hline
	  10 & 0.8247 & 4.52 & 0.8075 & 5.07 & 0.8533 & 5.42\\
	  20 & 0.8303 & 4.91 & 0.7928 & 4.93 & \bf{0.8550} & \bf{5.62}\\
	  50 & 0.8168 & 5.09 & 0.7527 & 4.31 & 0.8287 & 5.54\\
   	  100 & 0.7903 & 4.98 & 0.7037 & 3.72 & 0.7985 & 5.21\\
      200 & 0.7605 & 4.62 & 0.6539 & 3.11 & 0.7663 & 4.77\\
	  \hline\end{tabular}
	  \vspace*{-3mm}
  \end{table}

  \begin{table}[b!]
  \vspace*{-6mm}
  \caption{Varying hierarchical clustering structure by changing (a) the number of clusters per node at each level and (b) the number of levels on \texttt{Epinions} dataset. Results show that MSLP is not only more robust than CLRA to different clustering structures, but also outperforms other methods in most cases. Percentage is the percentage of within-cluster edges (Numbers in brackets represent percentage of within-cluster edges of random clustering).}
    \label{table:varying_cluster}
    \centering 
    \subtable[Changing the number of clusters per node at each level.]{
	  \begin{tabular}{c|c||r|r||r|r}	\hline
	  \multirow{2}{*}{$c$} & \multirow{2}{*}{Percentage} & \multicolumn{2}{|c||}{CLRA-Katz} & \multicolumn{2}{|c}{MSLP-Katz} \\ \cline{3-6}
	  & & AUC & Prec & AUC & Prec \\
	  \hline\hline
	  2 & 53.41 (13.42) & 0.7928 & 4.93 & \bf{0.8550} & \bf{5.62} \\
	  3 & 41.91 (6.41) & 0.7649 & 4.32 & 0.8520 & 5.48\\
   	  4 & 37.33 (5.47) & 0.7426 & 3.80 & 0.8463 & 5.27\\
      5 & 34.03 (3.12) & 0.7293 & 3.74 & 0.8276 & 4.80\\
	  \hline\end{tabular}
	  \label{table:ncs}}
	  \vspace*{-2mm}
	\subtable[Changing the number of levels.]{  
	  \begin{tabular}{c|c||r|r||r|r}	\hline
	  \multirow{2}{*}{$\ell$} & \multirow{2}{*}{Percentage} & \multicolumn{2}{|c||}{CLRA-Katz} & \multicolumn{2}{|c}{MSLP-Katz} \\ \cline{3-6}
	  & & AUC & Prec & AUC & Prec \\
	  \hline\hline
	  2 & 67.54 (26.51) & 0.7970 & 5.02 & 0.8459 & 5.33 \\
	  3 & 53.31 (13.42) & 0.7928 & 4.93 & \bf{0.8550} & \bf{5.62}\\
   	  4 & 47.77 (8.59) & 0.7825 & 4.60 & 0.8508 & 5.55\\
      5 & 43.54 (5.39) & 0.7633 & 4.15 & 0.8498 & 5.37\\
	  \hline\end{tabular}
	  \label{table:lvl}}
	  \vspace*{-2mm}
	 \subtable[Results of other methods.]{  
	  \begin{tabular}{l||r|r}	\hline
	  Method & AUC & Prec \\
	  \hline\hline
	  PA(Preferential Attachment) & 0.7717 & 2.09 \\
	  AA(Adamic-Adar) & 0.8378 & 5.16 \\
      RWR(Random Walk /w Restarts) & 0.8468 & 2.68 \\
      LR(Logistic Regression) & 0.8227 & 4.60 \\
      CN(Common Neighbors) & 0.8163 & 4.78 \\
      Katz(Katz) & 0.8352 & 4.77 \\
	  \hline\end{tabular}
	  \label{table:epin_other}}
	  
  \end{table}

\vspace{1ex}
\noindent {\bf Hierarchical clustering structure:}
Next we experiment with various hierarchical clustering structures. Table \ref{table:varying_cluster} shows how the performance changes as the hierarchical clustering structure changes. For a complete comparison, results of other methods are also given in Table \ref{table:epin_other}. The second column in Tables \ref{table:ncs} and \ref{table:lvl} represents the percentage of within-cluster edges. It is clear that as the number of clusters at the bottom level increases the percentage decreases. While the accuracy of CLRA degrades as the percentage decreases, MSLP is still able to perform better than other methods in all cases with the only exception of Table \ref{table:ncs} at $c=5$. The results clearly show that MSLP is robust to different hierarchical structures.

\begin{figure}[b!]
\centering
\includegraphics[scale=0.35]{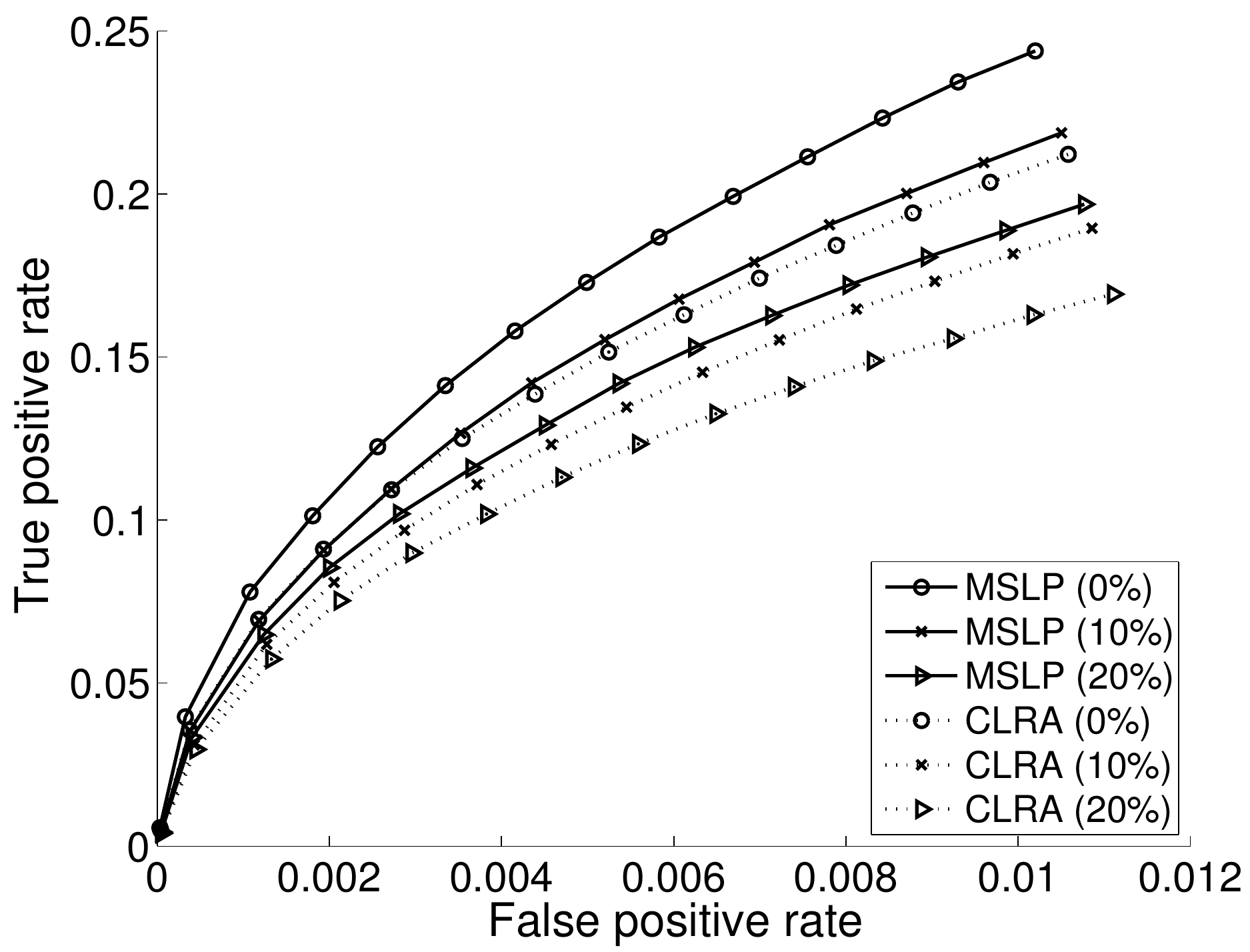}
\vspace*{-3mm}
\caption{Robustness results for MSLP and CLRA on \texttt{Epinions} dataset. Numbers in brackets are the \% of vertices shuffled.}
\label{fig:robust}
\end{figure}

Furthermore, to see that MSLP is robust to poor cluster structures, we randomly shuffle clusters in the deepest level of the hierarchy by moving vertices from their original cluster to another random cluster. Figure \ref{fig:robust} shows the result of moving 0\%, 10\% and 20\% of vertices. Even with 10\% of vertices shuffled, MSLP still outperforms CLRA with no shuffling. It is clear that, while CLRA's performance decreases rapidly, MSLP still performs well in low FPR regions.

\subsubsection{Results on Large-scale social networks}

  \begin{table}[b!]
	  \caption{AUC results for \texttt{Flickr}, \texttt{LiveJournal} and \texttt{MySpace} datasets.}
	  \label{table:auc}
	\centering
	  \begin{tabular}{l||r|r|r}	\hline
	  Method & \texttt{Flickr} & \texttt{LiveJournal} & \texttt{MySpace} \\
	  \hline\hline
	  PA & 0.6981 & 0.6739 & 0.8325 \\
	  AA & 0.8758 & 0.8316 & 0.8881\\
	  RWR & 0.8364 & 0.7972 & 0.8357\\
	  LR & 0.7784 & 0.7816 & 0.8359 \\ \hline
	  CN & 0.8649 & 0.7657 & 0.8819\\
  	  EIG-CN & 0.8759 & 0.7777 & 0.8764\\
	  CLRA-CN & 0.8697 & 0.7936 & 0.8669\\
	  MSLP-CN & 0.8783 & 0.8077 & 0.8788\\ \hline
	  Katz & 0.8634 & 0.8353 & 0.8673\\
	  EIG-Katz & 0.8888 & 0.8012 & 0.8823\\
	  CLRA-Katz & 0.8768 & 0.8056 & 0.8803\\
	  MSLP-Katz & \bf{0.9097} & \bf{0.8414} & \bf{0.8996}\\
	  \hline\end{tabular}
  \end{table}

  \begin{table}[b!]
	  \caption{Precision at top-100 results for \texttt{Flickr}, \texttt{LiveJournal} and \texttt{MySpace} datasets.}
	  \label{table:prec100}
	\centering
	  \begin{tabular}{l||r|r|r}	\hline
	  Method & \texttt{Flickr} & \texttt{LiveJournal} & \texttt{MySpace} \\
	  \hline\hline
	  PA & 1.02 & 1.32 & 4.57 \\
	  AA & 7.29 & 5.93 & 7.44\\
	  RWR & 5.49 & 3.46 & 1.30\\
	  LR & 2.54 & 2.23 & 4.95 \\ \hline
	  CN & 7.08 & 5.94 & 7.18\\
  	  EIG-CN & 6.88 & 5.34 & 6.99\\
	  CLRA-CN & 6.91 & 5.21 & 6.88\\
	  MSLP-CN & 7.03 & 5.59 & 7.05\\ \hline
	  Katz & 7.17 & 5.86 & 6.18\\
	  EIG-Katz & 11.26 & 5.62 & 7.55\\
	  CLRA-Katz & 12.13 & 6.11 & 7.64\\
	  MSLP-Katz & \bf{13.34} & \bf{6.72} & \bf{8.83}\\
	  \hline\end{tabular}
  \end{table}
  
In this section, we present results on real-world networks with millions of nodes presented in Table \ref{table:dataset2}. We construct a hierarchical structure with $\ell=5$ and $c=2$ for all three networks, and use $r=100$ for EIG, CLRA and MSLP. Tables \ref{table:auc} and \ref{table:prec100} give AUC and precision at top-100 results for the various methods, respectively. MSLP-Katz gives a significant improvement over the Katz measure and outperforms all other methods. Specifically, it gains a relative improvement of up to 4\% in AUC and 15\% in precision over the next best performing method. In contrast, MSLP-CN remains comparable to CN, but performs better than EIG and CLRA. Superisingly, the supervised method LR does not perform well, which is consistant with results found in \cite{Backstrom:2011bi}. Note that we only use network-based features and no additional features for training. However, engineering for more features is a difficult task and constructing good features itself is computationally expensive.

\begin{figure*}[t!]\centering
 	\subfigure[\texttt{Flickr}]{\includegraphics[scale=0.28]{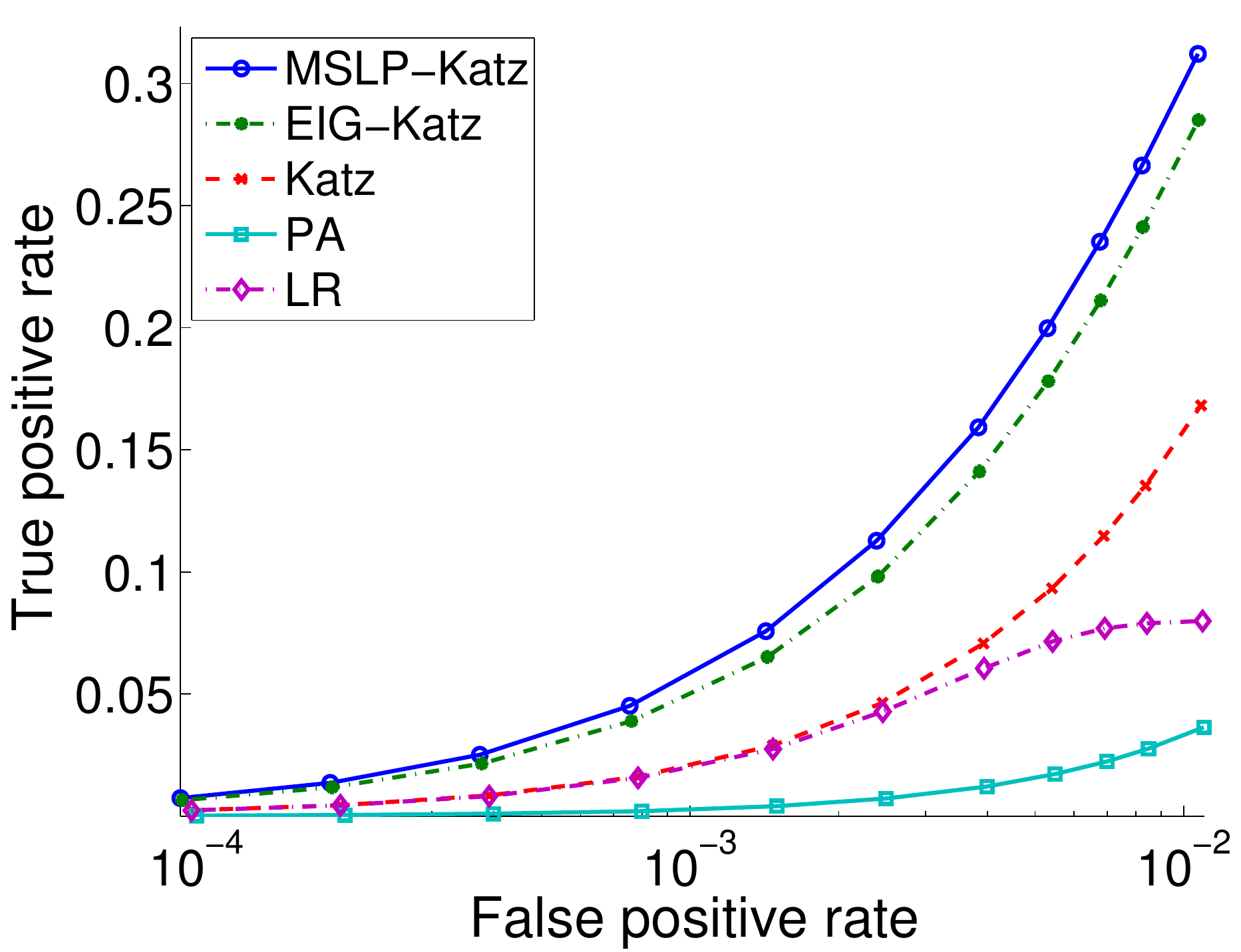}
 	\label{fig:fl}}
 	\subfigure[\texttt{LiveJournal}]{\includegraphics[scale=0.28]{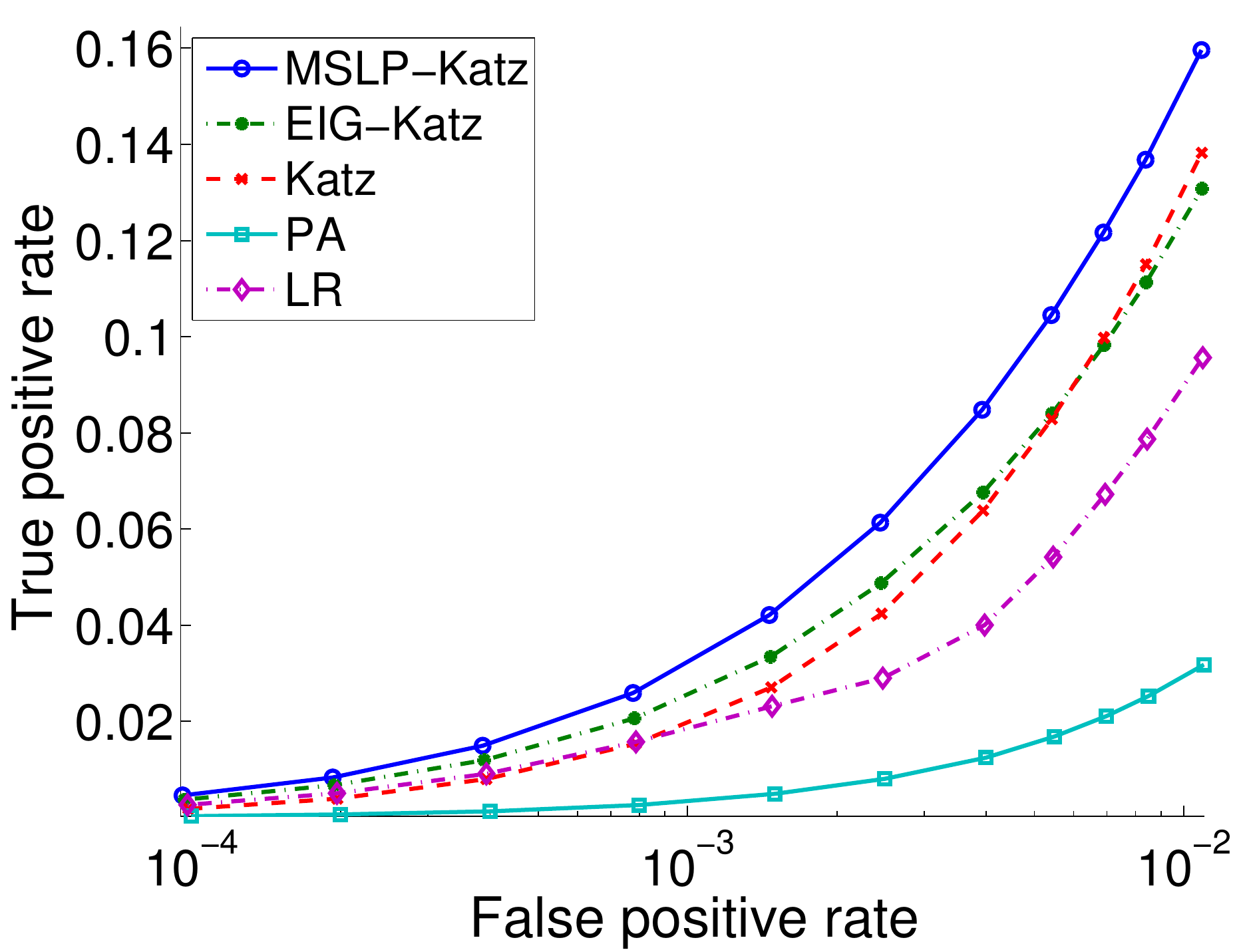}
 	\label{fig:lj}}
 	\subfigure[\texttt{MySpace}]{\includegraphics[scale=0.28]{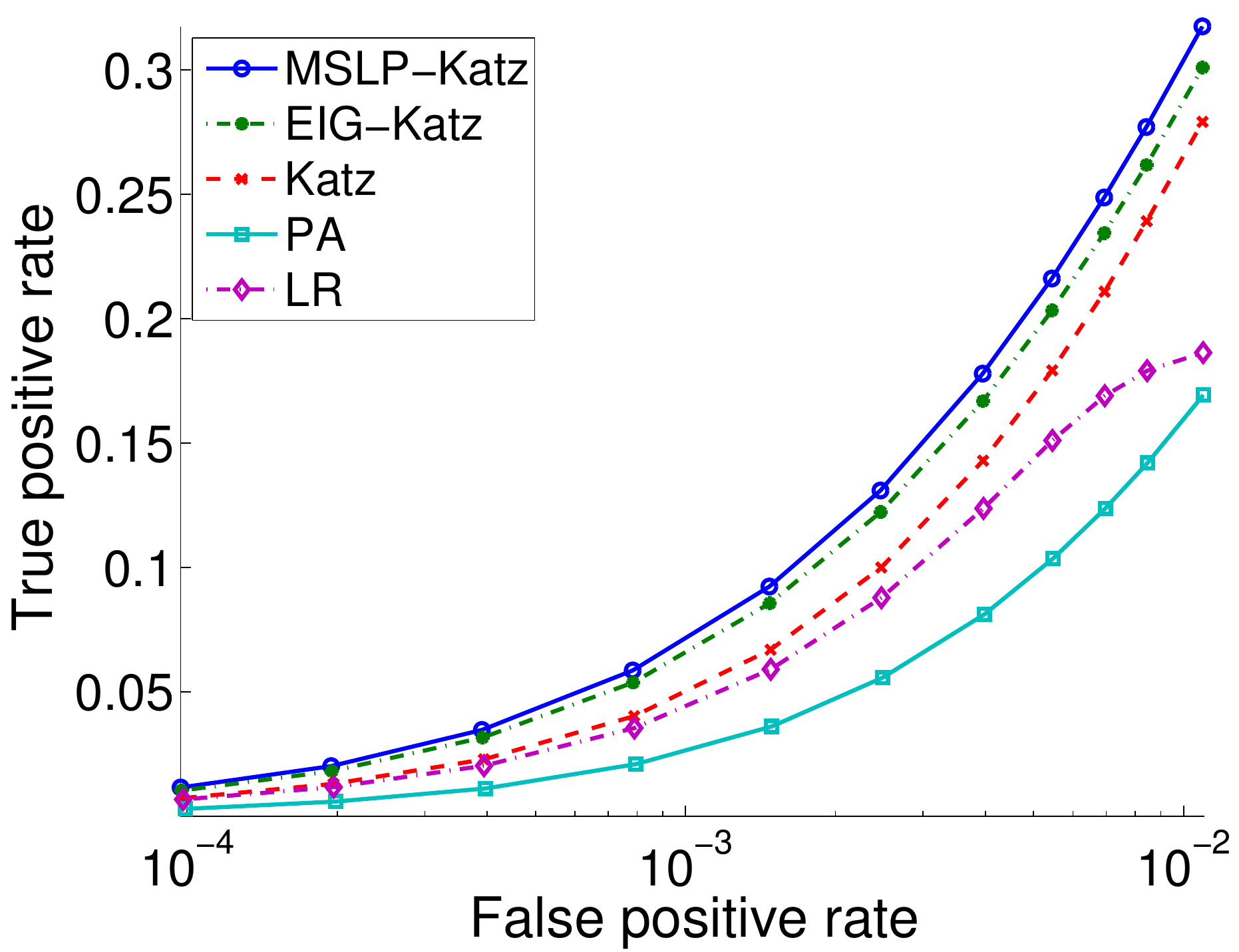}
 	\label{fig:ms}}
 	\caption{ROC curve on low FPR region of different methods for \texttt{Flickr}, \texttt{LiveJournal} and \texttt{MySpace} datasets. MSLP performs the best in all three datasets.}
 	\label{fig:link_prediction_results}
   \end{figure*}

\begin{figure}[t!]
 \centering
 \includegraphics[scale=0.4]{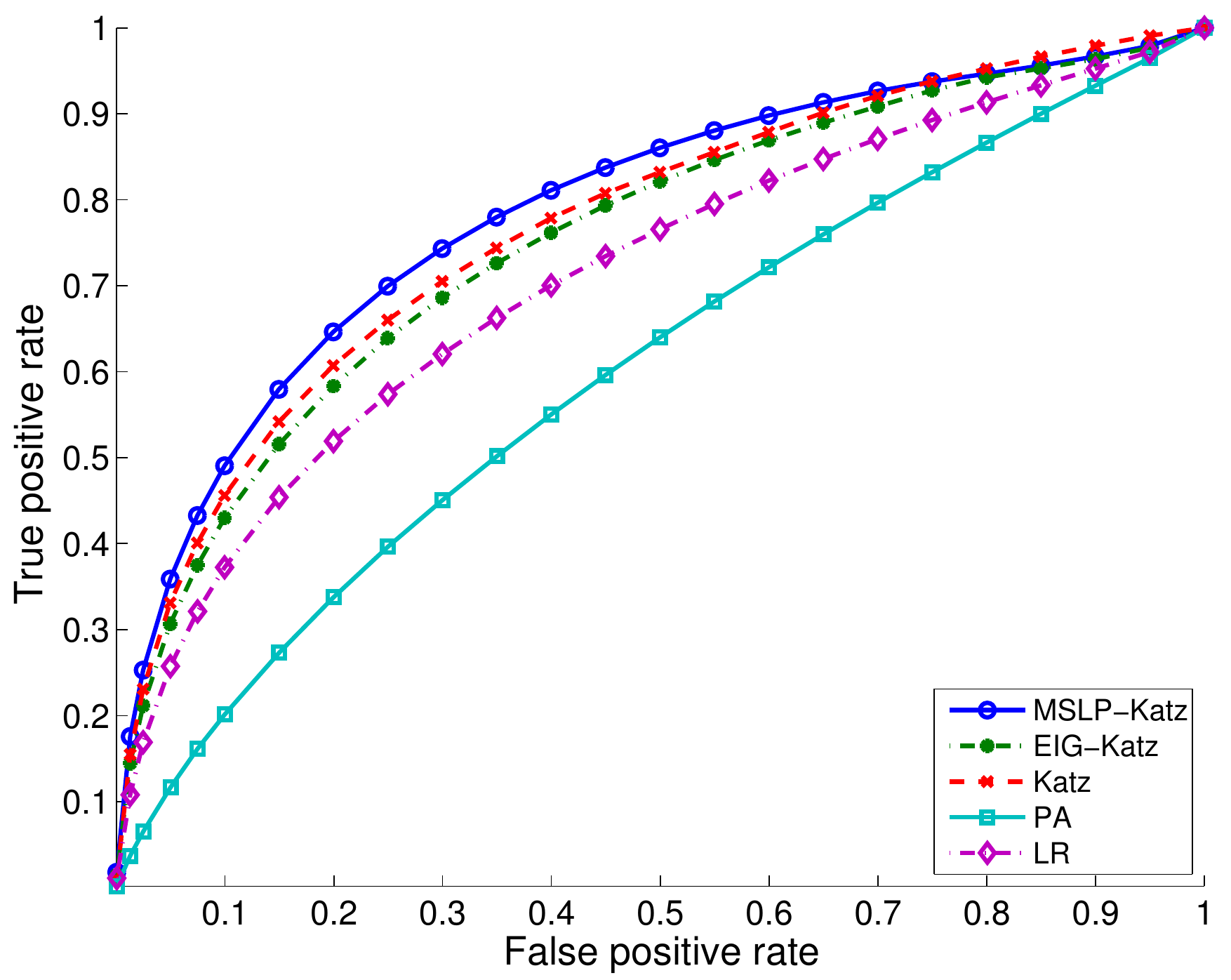}
 \caption{Full ROC curve for \texttt{LiveJournal}.}
 \label{fig:lj-hop2-full-roc}
 \vspace*{-3mm}
 \end{figure}

Figure \ref{fig:link_prediction_results} gives ROC curves focused on the low FPR region for the three large-scale networks. We note that only one representative method from methods that have similar performance is plotted for the sake of clarity. We observe that MSLP-Katz performs the best in all three datasets with significant improvements over Katz. For a given TPR, MSLP reduces FPR by $10\%$ on average and at most $20\%$ compared to others in all datasets. For completeness, we present the full range of the ROC curve for \texttt{LiveJournal} in Figure \ref{fig:lj-hop2-full-roc}. Note that much of the performance boost comes from the left side of the curve, which corresponds to the area of interest. That is, MSLP achieves good prediction quality for the highest predicted scores.

While dimensionality reduction methods, such as EIG and CLRA, tend to perform well in all three datasets, they are limited to a single low-rank representation of the network. Furthermore, CLRA has the largest drop in relative performance in terms of precision compared to MSLP in the \texttt{MySpace} dataset, where only 56\% of the edges are within clusters, whereas MSLP achieves the best result. Overall, the superior performance of MSLP illustrates the effectiveness of our multi-scale approach.

We note that the majority of time is taken by computing CLRA at the deepest level and thereafter low-rank approximations of upper levels can be obtained efficiently due to Algorithm \ref{alg:alg1}. However, CLRA can be easily parallelized as computing the subspace of each cluster is independent to other clusters. Thus, MSLP can achieve much more speedup with such implementation and serve as a highly scalable method for link prediction.

\section{Conclusions}
In this paper, we have presented a general framework for multi-scale link prediction by combining predictions from 
multiple scales using hierarchical clustering. A novel tree-structured approximation method is proposed to achieve fast and scalable multi-scale approximations.
Extensive experimental results on large real-world datasets have been presented to demonstrate the effectiveness of our method.
This significantly widens the accessibility of state-of-the-art proximity measures for large-scale 
applications. 

For future work, we plan to investigate methods to learn the weights for all 
levels following the hierarchy. It makes sense since some levels of the 
hierarchy have better predictions and deserve larger weights in the final 
prediction. This is also related to the hierarchical clustering structure we are using.
In this work, we use a balanced hierarchical structure mainly for its simplicity in combining predictions. However, a more realistic setting would be to use an unbalanced hierarchical clustering structure. Here, the issue is how to combine predictions from different levels as some links may not receive predictions at certain levels.
We also plan to parallelize MSLP as it can be parallelized within 
each level of the hierarchy and deal with the unbalanced hierarchy. 

\section{Acknowledgments}
This research was supported by NSF grants CCF-1117055 and CCF-0916309.

\bibliographystyle{abbrv}
\bibliography{arxiv-mslp}  

\end{document}